# A project-based course about outreach in a physics curriculum

Julien Bobroff, Frédéric Bouquet
*Laboratoire de Physique des Solides, CNRS, Univ. Paris-Sud, Université Paris-Saclay, 91405 Orsay, France* (Date : November 5, 2015)
E-mail : julien.bobroff@u-psud.fr

**Abstract**

We describe an undergraduate course where physics students are asked to conceive an outreach project of their own. This project-based-learning course alternates between the project conception and teaching activities about outreach. It ends in a public show. Students decide the topic and format on their own. An analysis of the students' productions over three years shows that all physics fields were equally covered, and various formats were used (experimental devices, animation or fiction movies, games, live events, photography). Some typical examples are described. We also analyze the benefits of this approach from the students' perspective, through a survey done over three classes. Students showed an overall very good assessment of the course (average of 4.5(0.6) on an appreciation scale from 1 to 5) and recognized having developed outreach skills but also project-management and group-work know-how. They acknowledged this course to be a unique opportunity to share with an audience their interest in physics compared to other courses. They further mentioned that it served as an intermission in a classical academic curriculum. They also point out some challenges, especially the time-consuming issue. This survey together with the practical description of the course implementation should help other universities develop similar courses.

## 1. Introduction

Physics students have various ways to learn about outreach and the communication of science. Physics curricula sometimes include courses that offer a theoretical approach to topics that involve science of communication, sociology, psychology or didactics [1,2]. Courses can also be more hands-on, where students are asked to develop and undertake a specific outreach action in front of an audience, such as hands-on demonstrations at a science museum [3], or lay-language presentations to the general public [4]. But more often, students discover outreach through their own endeavors, in informal science contexts. A common example is students conceiving together a show with live demonstrations of physics phenomena for the general public, given in a theatre, a science museum or a science fair in a sharing spirit. In Europe, forty-seven student groups have been identified who give such shows [5].

Both approaches have advantages: lectured-based course can provide structure and grounding while physics shows and out-of-the-class endeavors capitalize on personal engagement. This is why we chose to develop an outreach course integrated in a pedagogical project-based framework. Project-based learning has indeed been shown to foster among students teamwork, engagement and communication skills [7,8], qualities that are well suited for an outreach activity. In this course given at the third year undergraduate level in a physics curriculum, students make up their own outreach project and are encouraged to explore original formats. At the end of the course, they have to collectively produce a physics show for the general public.

The objectives of the course are to develop communication and didactical skills, to use project-based teaching to develop learning-by-doing, team work, project management and interdisciplinary skills [7], to open students to new forms of outreach and to question their role as scientists in society [9,10], and to offer the students a new approach of physics.

In this article, we describe the course and the projects conceived by the students over the past three years, together with a survey on the students' opinion about the course. Our main motivations are 1) to offer a detailed description of the course so that it can be easily duplicated in other universities and 2) to identify the benefits from the student's perspective.

## 2. The course description

A project-based learning is built on a question or problem that drives the learning activities of the students in designing a final product that addresses that central challenge [7,11]. In this course, the driving problem is: how to engage the public with the fundamentals of physics? Students are asked to design a concrete physics outreach artifact or activity to test possible solutions to this problem. This course is one of the electives proposed to third year undergraduates in the "fundamental physics" section at the University of Paris-Sud. This section is considered in France as one of the most demanding in physics in terms of level and schedule (thirty to thirty-five class hours per week, most being formal courses on modern physics topics). Electives are proposed to students, offering an opening to domains not covered by the majors, such as astrophysics, biophysics, geophysics or history of science. The outreach course has been proposed as such an elective for the past three years, with a maximum limit of about twenty students for practical reasons. Each year, about fifteen to twenty students representing 15% of the full population select this elective among a choice of eight. Passing this course brings the students 4 European Credits, on the 30 needed to validate the semester.

The course consists of thirteen sessions of three hours each, given weekly over a semester. In addition, students spend typically one to three hours weekly working on the project out of class. Two teachers (authors of this article) mentor the students organized into one or two classrooms depending on the teaching sequence. The teachers are experienced in outreach, having participated or organized science fairs, outreach talks, and collaborated with science museums [12]. The semester is divided in three phases: two starting sessions with very small projects, nine sessions devoted to a large project, and two sessions to prepare a public show where students present their projects to an audience. All through these three phases, accompanying courses and exercises are proposed related to practical or more conceptual outreach aspects.

In the first two-session phase, students are asked to achieve a hands-on specific challenge in a limited amount of time under various constraints. For example, after a short introduction on how to handle a superconducting levitation experiment, students are asked to produce in two hours a short video of an experiment satisfying two randomly picked constraints, one about the style of the video (funny, scary, didactic…), the other about the use of an additional element (aluminum, plastic glass, laser pointer, balloon…). In another of these short challenges, the students have to prepare and manipulate conducting and insulating play dough to create original electrical circuits [13]. These starting challenges aim at immediately engaging students in learning-by-doing activities and building a team spirit. They also help demonstrate to the students that outreach can embody various and sometimes unexpected formats. These formats often play at odds with their preconceptions about outreach consisting only in didactic talks to a sitting audience.

In the second and longest nine-session phase, students are asked to conceive an outreach project in groups of two to three. Teachers do not propose a list of possible projects. Instead, a collective brainstorming session is organized among students where they have to imagine possible topics, formats, and outputs which then allows them to decide what to work on. Their choice is based on personal preference, but also expertise, skills and interpersonal relations with other students. This freedom of choice with no suggested topics is a key ingredient to engage the students with a challenge that they have now taken ownership of. A presentation of various examples of recent original communication formats is also given before this brainstorming to push students to explore non-traditional routes. Then, students have to schedule and manage all the aspects of their project during the nine sessions: bibliography, purchasing of materials, conception, testing, etc. They are asked to produce at the end an operational artifact that can be used and presented to the public, not just an on-going project. This requirement is a major constraint for them.

In the last two sessions, students are asked to design and rehearse a consistent and appealing one-hour physics show. They have to elaborate the staging including technical aspects (lights, music, announcers) and imagine ways to make the show more than just a mere juxtaposition of their projects. This show is given at the end of the semester to a typical audience of about 200 to 300 university students, researchers or administrative staff. This public show is the essence of what outreach is about, i.e., meeting with the public. We noticed that the show helped induce a team spirit within and between the groups. Furthermore, its prospect pushes the students to reassess and often scope down their initial goals during the course. On downside, some students reported that the show had been stressful to them.

In addition to the project conception, the teachers give six half-hour to one hour presentations followed by practical exercises on various facets of outreach: how to prepare an outreach slideshow; how to explain a formula to the general public; how to explain a physics concept under a given constraint (mime, drawing, dance); how to edit an outreach video; how to use Arduino open-source technologies, exploring the new formats of outreach (such as YouTube).

The students' final marks are determined by the teachers based on their experience of outreach activities. Three criteria are used: students' involvement and participation evaluated all along the course (40% of the final mark), the quality of the project itself (40%) and the students step-back (20%). The latter criterion is judged through a final oral exam, during which students present their motivations, a study about similar other existing projects, their own project with a critical point of view and possible perspectives. The fact that grades are attributed by teachers only is perhaps not optimal since a good project-based evaluation may also involve students self-evaluation and the recipient of the project (here the show's audience) [11].

On practical aspects, a need for at least two classrooms has proven essential since students require room and often silence for shooting videos, preparing complex experiments or just rehearsing oral sequences. As for equipment, a set of multimedia tools is required (video cameras, cameras, tripods, external lightings, microphones, laptops with free video and audio editing tools), all of which corresponds to about $3000 in initial budget. Basic lab equipment available in physics departments is also useful. Specific low cost equipment can be required after the projects' topics are decided, so a modest additional flexible budget should be anticipated.

### 3. The students projects

#### 3.1. Subjects and formats

We analyzed all the students' projects over the past three years, which all can be consulted in [14]. Over the twenty-three projects conceived, the topics are found to cover all the fields of physics. No preferred topic was found, contrary to the fact that most students come to this specific physics section based on a motivation toward astrophysics or quantum physics. 30% of the projects deal with topics already presented in high school (mechanics, acoustics, electromagnetism) while 58% relate to undergraduate physics fields (quantum physics, relativity, thermodynamics, particle physics, nanophysics, solid state physics, astrophysics). 12% relate to history, epistemology or sociology of science.

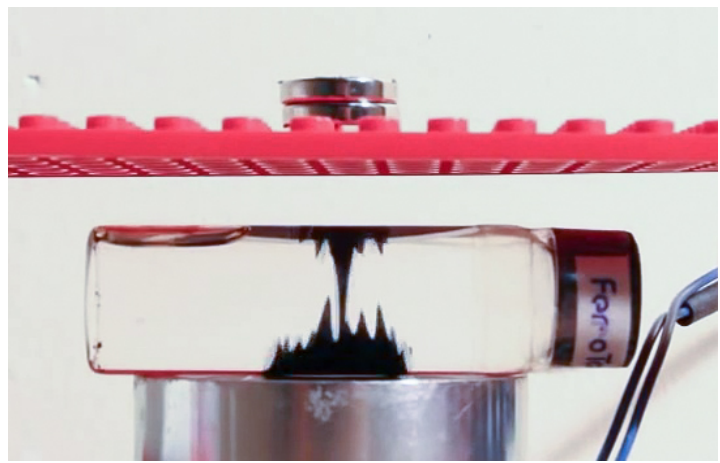

**Figure 1**. "Make the magnets dance": a ferrofluid (black liquid) is placed below permanent magnets and on top of an electromagnet. When music is played through the electromagnet, the ferrofluid dances in rhythm.

Various formats were explored: experimental devices (39%), animation movies (22%), fiction movies (17%), games (9%), live events (9%), and photography (4%). This diversity is a natural consequence of the brainstorming session approach and is one of the known benefits of project-based learning [11,15]. It also reflects the recent evolution of popularization of science into various new media and artifacts: short videos on internet, theatre, dance, comics, songs, art, geek and pop culture, happenings, citizen science, etc. [6]. Here are some examples of the projects, which we have chosen to be representative of the various formats encountered.

"See the invisible" is a didactic experimental device about Magnetic Force Microscopy (MFM). It displays a human-scale model of the way a MFM works: a laser pointer beam is reflected on a small mirror attached to a magnetic tip. The tip is positioned on the top of a large-scale model of a credit card. The magnetic stripe of the card consists of hidden NdFeB flat magnets facing north or south randomly. When displaced over the magnets, the magnetic tip is attracted or repelled and the mirror deflects the laser beam. The observation of the laser deflection on a wall a few meters from the setup allows a determination of the magnets' orientations, and therefore to 'read' the magnetic strip. This basic device mimics the actual operation of a MFM but magnified to our scale. The students also worked on an explanation talk about real MFM set-ups and magnetic imaging which accompanied the setup and live demonstration during the show.

"Make the magnets dance" is also an experimental setup about magnetism, but more artistic and contemplative. A ferrofluid floating into a water tube is positioned between a static magnet and an electromagnet, as displayed in figure 1. When the music is played through the speaker and electromagnet, it creates a magnetic field, which induces forces on the ferrofluid. These forces make the ferrofluid 'dance' in rhythm with the music. Here again, the live demonstration was accompanied by an additional slideshow to explain the physics at play.

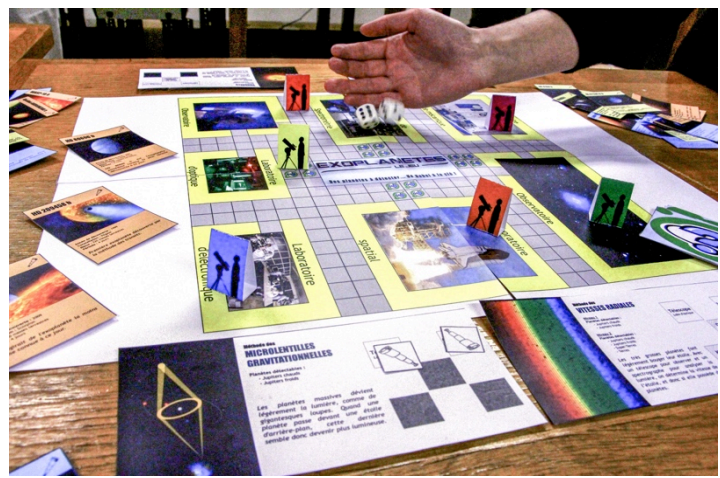

**Figure 2.** A board game where gamers try to win the Nobel Price by discovering new exoplanets using labs and observatories.

"Women physicists" consists of an animated movie portraying the life of Cecilia Payne, a twentieth century astrophysicist who explained the composition of stars. Her life is depicted in a 'Draw My Life' style, using fast-motion photography of handmade drawings on a white board. These types of stop-motion animations have become popular on YouTube, and even in physics outreach, as in the Minute Physics video channel [16]. The video is followed by three similar animations about the physics itself (eclipses, star formation and spectrometry). Students played a small theatrical scene during the show to introduce the video and engage with the audience about gender issues in physics.

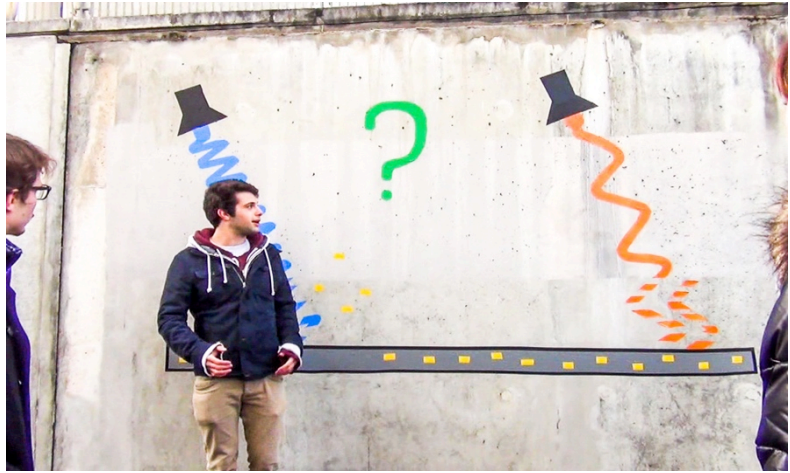

**Figure 3.** A student explains the photo-electric effect using tape drawings made on a wall in the city.

Some projects explored more tangible non-digital formats. For example "the Exoplanet game" is a board game about discoveries of new planets out of the solar system. In order to win the Nobel Price, players have to discover various planets using labs and observatories. They navigate in a Cluedo-like board game as seen in figure 2 and answer various questions about astrophysics, Sci-Fi movies, or history of science. The game is available for download online and can be assembled with just paper and glue. "Tape light" is based on another unusual format: tape art. The students went in the city and used colored scotch tapes to draw graffiti on a wall to present the photoelectric effect as in figure 3. It allowed them to engage in a dialog with passers-by about quantum physics and its applications.

### 3.2. Analysis of the projects

Over the past three years, none of the twenty-three projects undertaken by the students followed any of the traditional formats of physics outreach (talks, posters, articles). This outcome is probably the result of the way we presented outreach and of the brainstorming setup (see previous section). Projects were taken as an opportunity for students to use new tools and formats they never use in the French physics curriculum. It also allowed them to incorporate inputs from their own cultural background out of the academic world. For example, references to TV or Internet series were often claimed, such as "CSI: Crime Scene Investigation" or the French YouTube series "Les Visiteurs". Geek and pop-culture were also influential, as demonstrated by the use of Playmobil, Legos or tape art. We were surprised that almost no project involved the use of coding or numerical interface or high-tech digital tools such as Kinect or Arduino technologies. These choices could be related to the comparatively low-skill level in computer science among this student population.

Finally, we – as teachers – sometimes had the feeling that the physics itself – even though present in all projects – was superficially treated. For example, in the project "quantum crime", a detective movie about quantum physics, students based their scenario on entanglement and quantum measurement, but their understanding remained superficial.

### 4. Assessing the course through a student survey

We focus in this study on the students' perception of the course, which we assessed through a survey sent to all participants to this outreach course over the past three years. We deliberately did not choose to assess their actual outreach skills. This would require a survey carried out on the audience of the final shows in various environments (schools, general public…), which is out of our scope here. Objective assessment of project-based curricula is hard to conduct [11]. Ideally, a long-term survey should be carried out by outside educational researchers on all the students comparing them to a test-population of similar students who did not participate in the course. Instead, our assessment methodology induces various biases: 1) the students who agreed to answer may be more favorable to this teaching approach than the others, 2) students answered to their former teachers 3) the teachers carried out the analysis themselves. Taking those provisos into account, here are the main features of our analysis of the survey.

Over the fifty-two students who were sent the survey, twenty-five answered. Contrary to our expectation, the students who took the course more recently answered less frequently: 67% of the students who followed the course three years before the survey answered, to be compared to 53% for two years and 40% for one year. This difference may signal that the more students progress in their studies, the more valuable they consider this course's role to be in their curriculum.

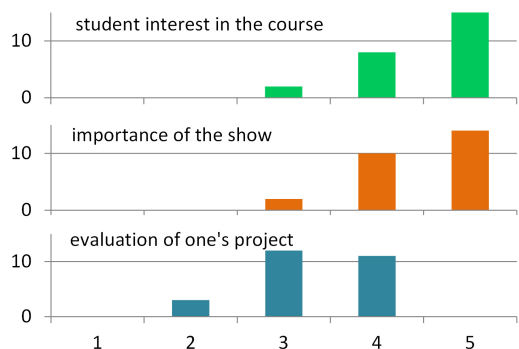

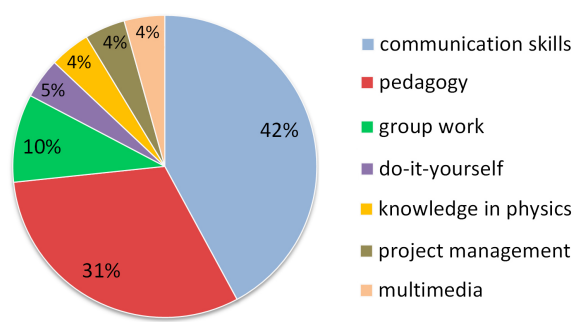

**Figure 4.** student assessment of the course (1: no interest – 5 : very interesting), of the importance of the show (1: no interest – 5 : essential), and of their own projects (1: very bad – 5 : very good).

**Figure 5.** student answers to a multiple choice question "which skill do you think you learned the most that you did not have before?" (note that "modern research" and "history of science" were also proposed but never chosen).

Fig. 4 shows the students' opinions about the course, the show, and their own projects. Students assessed their interest in the course at an average of 4.5(0.6) on a Likert scale from 1 to 5 [17]. When asked if they would have preferred to choose another elective instead, 96% answered 'no' and 4% 'without opinion'. They consider the show to be very important, as demonstrated by an average score of 4.5(0.6) (1: no interest – 5: essential), confirming that the show plays a major role in their motivation. As their projects are concerned, students appear more critical since they evaluated their own project with an average of 3.4(0.6). Figure 5 shows the answers to the closed-ended question "which skill do you think you learned the most that you did not have before ?". Students clearly favored oral and communication skills to more academic or technical skills. This opinion might be reinforced by the fact that students compared this course to their more classical physics courses mostly focused on content.

Students were asked an open question on how they experienced that course compared to the rest of the curriculum and their life at the time. 39% claimed that the course served as an area of freedom in the otherwise formal curriculum. Students' quotes: "it was an intermission between two theoretical courses", "a big breathing [space] in the intense schedule", "refreshing because you actually do, see and talk about physics differently". 22% pointed out it was original: "this teaching is really different because of the personal involvement", "Topics I had to work on were also really different than other courses", "there was a strong contrast with other courses (which are) more academic". 22% complained that it was too time-consuming and would have required more time to do well. 13% argued it proposed a new way to deal with physics: "I appreciated (the chance) to lead projects with autonomy where one could both take one's mind off things, be creative, but be anchored in a science background".

In some of the comments, students who did not rank well in the other physics courses emphasized that this course was sort of a sanctuary: "as I was having poor grades at that time, the outreach course served as an (…) intermission where I did not have the feeling of being behind the other students in understanding the content". Isolated answers mentioned the course helped to develop new skills and creativity, to be playful, and one student mentions, on the negative side, that he felt other students often considered this elective to not to be as academically serious as other courses.

In a final open section for "comments, criticisms and suggestions", 30% of the students took the opportunity to further praise the course, 17% suggested it deserved more hours, and 17% asked for more small pedagogical exercises.

## 5. Discussion

The main motivation of our study was to identify the benefits of this course from the student perspective. From the survey, students clearly experienced an enjoyable course and acknowledged developing specific skills: communication and teaching, group work, project management, do-it-yourself ability. A former student who is now a high school teacher commented, "the course helped (him) with new ideas and concepts which are now helpful for teaching and communication". Other former students now in PhD mentioned that they currently participate in outreach actions or initiate some in their labs. Moreover, students often told us that this course was their first opportunity to share with others, especially during the public show, an altruistic dimension usually absent from physics studies. Putting videos of the projects on Internet was also part of that sharing process [14]. This facet of the course also puts focus on their role as scientists in society [10]. Students could for the first time use personal skills in their physics studies, another source of pride and motivation. The survey also revealed that this course served as sort of a breathing space in an otherwise pretty intensive physics curriculum. Such a reprieve can be very beneficial, especially for students experiencing difficulties with the core curriculum.

We, as teachers, also found benefits in this course: the relationship we developed with students during the course was unique in terms of engagement, motivation, trust and even friendship compared to our other teaching experiences. It also helped us change our thinking about classroom structures, activities and tasks, which then influenced our other teachings, an influence also seen in other similar courses [7]. A final gain from the University perspective is that this course generates various media, which can be used both for outreach and for communication purposes at the university level to attract new students to physics sections.

We also encountered challenges. The course could be too time-consuming for students as they often became so focused on their projects that it became a detriment to their other courses. The time issue is the most often reported problem in project-based learning [11]. Second, we sometimes had to face interpersonal conflicts in teamwork, another common problem in project-based learning [18]. Another issue is that this course seems to scare away students with the top grades. The distribution of grades among students who took the course is similar to the total population, except for the top 10% students which is absent (only about 3% of these top students chose to take the course). Perhaps physics content in outreach teaching doesn't seem sufficiently rigorous for these profiles; and it is true that at least in some of the students' productions, physics at play was not deep enough. This is a known consequence of problem-based learning where a less rigorous understanding of the fundamentals is often encountered [19].

Overall, we consider that the benefits largely exceed the issues both from the students and teachers perspective. As the course is comparatively easy to implement and requires only one or two motivated physicists experienced in the field of outreach and a small initial equipment, we hope it will be developed in other universities and foster future physicists to be more engaged in outreach actions.

**Acknowledgements**

We gratefully acknowledge all the students who took part in this course and the survey, and Patrick Puzo and the Magistere de Physique who welcomed this original teaching in the curriculum. We also acknowledge Janet Rafner, Magali Fuchs-Gallezot and Laurence Maurines for their help with the manuscript. This work benefited from the support of the Chair "Physics Reimagined" led by Paris-Sud University and sponsored by Air liquide.